\documentclass[aps,prl,twocolumn,titlepage,footinbib,longbibliography]{revtex4-1}
\usepackage{amsmath,amssymb}
\usepackage{bm}
\usepackage{epsfig}
\usepackage{natbib}
\usepackage{hyperref}
\usepackage{color}
\usepackage{graphicx}
\usepackage{colortbl}

 \hypersetup{pdfstartview={FitH},pdfpagemode={UseNone},
            colorlinks,linkcolor=red, citecolor=blue, urlcolor=blue,
            bookmarks=true, bookmarksopen=true, pdfnewwindow=true}

\newcommand{\rmi}{{\rm i}}

\newcommand{\e}{{\rm e}}

\graphicspath{{./figs/}}

\begin{document}
\title{
Optomechanical Kerker effect
}

\author{\firstname{A.~V.} \surname{Poshakinskiy}}
\email{poshakinskiy@mail.ioffe.ru}
\affiliation{Ioffe  Institute, St.~Petersburg 194021, Russia}
\author{\firstname{A.~N.} \surname{Poddubny}}
\affiliation{Ioffe  Institute, St.~Petersburg 194021, Russia}

\begin{abstract}
  Tunable directional scattering  is of paramount importance for operation of antennas,  routing of light, and design of topologically protected optical states. For visible light scattered on a nanoparticle the  directionality could be provided by the Kerker effect,  exploiting the interference of electric and magnetic dipole emission patterns. However, magnetic optical resonances  in  small sub-100-nm particles are relativistically weak. 
  Here, we predict  inelastic  scattering with the unexpectedly strong tunable directivity up to 5.25  driven by a trembling of   small particle without any magnetic resonance.  The proposed optomechanical Kerker effect originates from the vibration-induced multipole conversion. 
 We also put forward an optomechanical spin Hall effect, the  inelastic 
polarization-dependent directional scattering.  Our results   uncover an intrinsically  multipolar nature of the interaction between light and mechanical motion.
They apply to a variety of systems from cold atoms to two-dimensional materials to superconducting qubits and can be instructive to engineer chiral optomechanical coupling.
\end{abstract}

\maketitle

Scattering of light  manifests itself  in everyday life,  fundamental science and device applications~\cite{Mischenko2002}.  Elastic Rayleigh scattering  governs the blue color of sky and sea. Inelastic Raman scattering  is a workhorse of  sensors.
The ability to control the direction, frequency, and polarization of the scattered light is essential for  optical devices.
However, both the Rayleigh and Raman scattering usually have a symmetric emission pattern: the waves are symmetrically scattered in the two opposite directions, in particular, forward and backward~\cite{Leite1965,Damen1965}. The asymmetry can be induced if the particle that scatters light moves. Then, the Doppler effect leads to a difference between the incident and scattered light frequencies~\cite{Raman1922book}, which depends on the scattering angle in a highly asymmetric way. As first noted by C.V.~Raman himself~\cite{Raman1919},  it vanishes for the forward-scattered wave and reaches maximum for the back-scattered one. The scattering cross-section depends on the angle between the incident light propagation direction and particle velocity, enabling cooling of atomic gases in optical molasses~\cite{Hansch1975}.
Still, the asymmetry of the emission intensity pattern remains small unless the particle velocity becomes comparable to that of light, which is realized, e.g., for Compton scattering of X-rays~\cite{Compton1923}. 

A simpler mechanism to achieve strong scattering  directionality was proposed by M.~Kerker~\cite{Kerker:83}. 
Rather than using mechanical motion, it requires a scatterer to possess both electric  (ED) and magnetic dipole (MD) susceptibilities.
Since the electric field of  these two modes is of the opposite spatial parity, their interference enables directional forward or backward scattering depending on the relative phase~\cite{Mehta2006,Geffrin2012,Liu2018}.
Thus, implementation of the Kerker effect requires magnetic response of the same strength as the electric one. At optical frequencies this becomes challenging since magnetic dipole transitions are intrinsically relativistically weak~\cite{landau08}. A promising recently emerged  workaround is provided by submicron-size high-refractive-index dielectric nanoparticles~\cite{Kuznetsov2012,Kuznetsov2016,Limonov2017} hosting both magnetic and electric Mie   resonances.  For instance, Huygens metasurfaces of Si nanodisks that transmit light forward changing its phase without reflection open new avenues for wavefront control at the nanoscale~\cite{Staude2013,Ding2018}. 
However, optical Kerker effect for the particles smaller than the wavelength in the medium $\sim$100\,nm is still unfeasible.
%%%%%%%%%%%%%%%%%%%%%%%%%%%%%%%%%%%
\begin{figure}[b!]
  \includegraphics[width=.99\columnwidth]{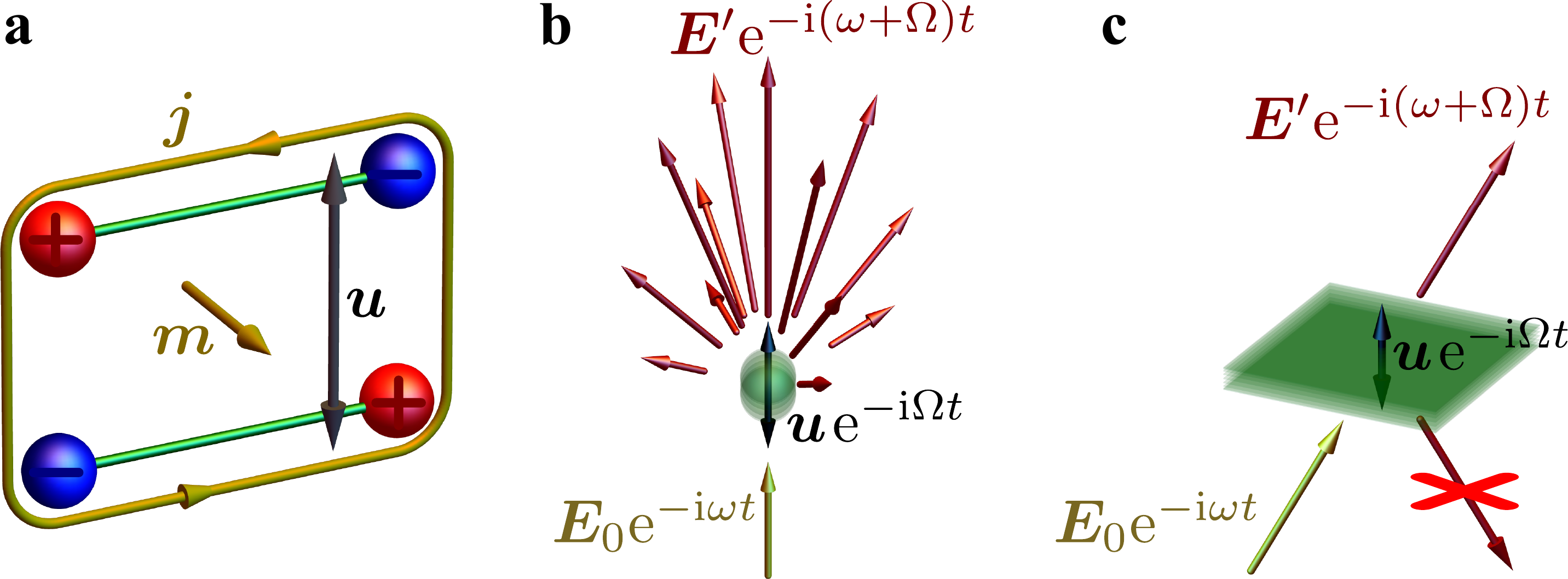}
\caption{{\bf Sketch of optomechanical Kerker effect.} (a) Trembling of an oscillating dipole in the space $\bm u$ leads to appearance of an electric quadrupole and a current curl. The latter yields a magnetic dipole $\bm m$. (b) Directional inelastic light scattering on a trembling particle. (c) A sketch of the trembling resonant layer, an optomechanical analogue of Huygens surface.}\label{fig:schem}
\end{figure}
%%%%%%%%%%%%%%%%%%%%%%%%%%%%%%%%%%%

Here, we uncover a deep nexus of the motion-induced scattering directionality and the Kerker effect. We  
 put forward an {\it optomechanical} Kerker effect, where strong tunable directionality is achieved for light scattered by a small particle without any magnetic response that trembles in space. 
% where the tunable directional forward or backward scattering is achieved for a small trembling particle without any magnetic response. 
Our main concept is sketched in Fig.~\ref{fig:schem}. The incident wave excites electric dipole polarization, that oscillates in time. Trembling of the electric dipole  in the direction transverse to its polarization  induces the loop electric current $\bm j$ with non-zero magnetic momentum $\bm m$ as well as the electric quadrupole (EQ) momentum.   Interference of ED and MD+EQ contributions results in unidirectional scattering   as shown in Fig.~\ref{fig:schem}(b,c).   
While the idea to use motion-induced conversion of electric dipole to magnetic dipole seems straightforward, 
a na\"{\i}ve expectation would be that the magnetic dipole is relativistically weaker than the electric one and their interference cannot result in any significant directionality. 
We found that magnetic and electric dipole components counterintuitively are of the same order when {\it inelastic} light scattering is considered. To demonstrate this, we have developed a  novel theoretical framework of multipolar resonant optomechanics. It  incorporates the effect of the resonant dispersion of the moving medium on the multipolar emission in a rigorous nonperturbative fashion and goes beyond previous approaches \cite{Padgett1998,Bulanov:2013,Kozlov2017}   restricted to non-resonant scatterers. Our predictions are quite general and  apply both for particles and for thin layers, as shown in  Figs.~\ref{fig:schem}(b) and~\ref{fig:schem}(c). We also put forward an {\it optomechanical spin Hall effect}, i.e. directional inelastic scattering of light depending on its circular polarization.

%%%%%%%%%%%%%%%%%%%%%%%%%%%%%%%%%%%%%%%%
\begin{figure}[tb]
  \includegraphics[width=.99\columnwidth]{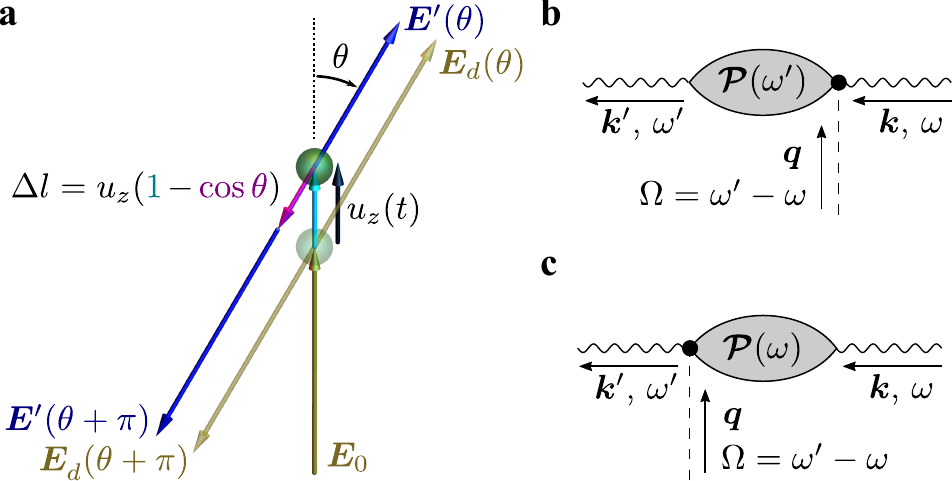}
\caption{{\bf Origin of directional inelastic scattering}. (a) A sketch of light scattering on a trembling particle. The incident and elastically scattered light are shown by yellow color, inelastically scattered light is shown by blue color. Inelastic scattering is caused by the temporal modulation of the optical path (cyan and magenta arrows) due to particle displacement.  (b) and (c) Diagrammatic representation for the inelastic light scattering on a trembling particle.  Wavy lines denote photons, bubbles correspond to the dressed polarization operator of the particle at rest, dashed lines represent mechanical displacement, solid dot stands for the optomechanical interaction given by Eq.~\eqref{eq:L} in Methods.}\label{fig:dtech}
\end{figure}
%%%%%%%%%%%%%%%%%%%%%%%%%%%%%%%%%%%%%%%%

%%%%%%%%%%%%%%%%%%%%%%%%%%%%%%%%%%%%%%%%
\section*{Results}

\subsection*{Directional inelastic backscattering}

We start with  the qualitative geometrical consideration to reveal a drastic difference in the angular patterns of elastic and inelastic scattering.  Figure~\ref{fig:dtech}(a) sketches the  plane wave  with the frequency $\omega$ that is scattered on a small particle trembling at the frequency $\Omega$ along the incident light propagation direction. The incident light induces dipole polarization of the particle that then emits light in a different direction. The shift of the particle in the real space $u_z(t)$ gives rise to an additional time-dependent  phase of the scattered light $\phi(\theta,t) =  (1-\cos\theta)(\omega/c) u_z(t)$, where the two terms are illustrated by cyan and magenta arrows in Fig.~\ref{fig:dtech}(a) and $\theta$ is the scattering angle.  The electric field of the scattered wave reads $\bm E'(\theta) = \bm E_d(\theta) \e^{\rmi\phi(\theta,t)}$ with $\bm E_d(\theta)$ being the electric field of the light scattered by the particle at rest.  Taking the particle displacement in the form $u_z(t) = u_z \e^{-\rmi\Omega t} + \text{c.c.}$ and expanding the scattered field into series over $u_z$, one obtains harmonics at the frequencies $\omega + p\Omega$ with integer $p$. 
We suppose that the vibration amplitude is small.
Then, the electric field of the harmonic at the initial light frequency $\omega$, that describes the elastic light scattering, coincides with $\bm E_d(\theta)$. Its angular dependence is governed by the well known electric dipole radiation pattern that yields equal amplitudes of forward and backward scattering~\cite{Novotny2006}. The  linear-in-$u_z$ terms yield the harmonics at anti-Stokes- and Stokes-shifted frequencies $\omega \pm \Omega$ that describe inelastic scattering. 
Their intensities  read 
\begin{align}\label{eq:intro}
I'(\theta) \approx (\omega/c)^2 |u_z|^2  (1-\cos\theta)^2 I_d(\theta) ,
\end{align}
where $I_d(\theta) \propto |E_d(\theta)|^2$ is the intensity of elastic dipole scattering. 
In stark contrast to elastic scattering, the  inelastic scattering is strongly anisotropic. In forwardscattering geometry, the particle shift does not change the optical path. Thus, Stokes and anti-Stokes light intensities vanish for $\theta=0$. The inelastic scattering is the most intensive in the backscattering geometry, $\theta=\pi$, when the optical path change is maximal.

%%%%%%%%%%%%%%%%%%%%%%%%%%%%%%%%%%%%%%%%
%\cite{Chen2011}

%\cite{Slobozhanyuk2016}
%\cite{Sounas2017} %nonreciprocal photonics based on time modulation
 %Time-reversal symmetry breaking with acoustic pumping of nanophotonic circuits

% We show that trembling particle with dipole resonance only manifests Kerker effect as well. That is because motion of and oscillating dipole leads to appearance of electric quadrupole (EQ) and magnetic dipole, see Fig.~\ref{fig:schem}.

\subsection{Multipolar resonant inelastic scattering}

The above geometric consideration  predicting the  inelastic scattering asymmetry has a crucial limitation: It is applicable only to the particle with the frequency-independent polarizability. Indeed, in the case of  resonant optical response, the elastic scattering  intensity $I_d$  strongly depends on the light frequency. Yet, it is completely unclear which frequency to choose in Eq.~\eqref{eq:intro}: either that of the incident light $\omega$ or that of the scattered light $\omega'= \omega\pm \Omega$. 

To resolve this fundamental problem, we develop a rigorous theory of light interaction with a  polarization of a trembling medium.  
We stress that the inelastic scattering considered here is caused by the motion-induced modulation of the interaction between light and scatterer, in contrast to conventional resonant Raman scattering that is due to modulation of the eigen energies of the scatterer itself.
We give the general expression for inelastic scattering intensity Eq.~\eqref{eq:S:S} in Methods, while here focus on the case of small trembling object described by 
 the frequency-dependent electric dipole polarizability tensor $\bm\alpha(\omega)$. 
  In that case, the amplitude of inelastic scattering comprises two terms that are 
diagrammatically represented in Fig.~\ref{fig:dtech}(b) and~(c). They show the vibration quantum (dashed line) being absorbed/emitted either before or after the medium polarization (bubble) is induced. This reflects the change of the optical path before and after the scattering on the particle, see cyan and magenta arrows in Fig.~\ref{fig:dtech}(a), respectively. Concomitantly, the two terms in the inelastic scattering amplitude feature polarization operators $\bm{\mathcal{P}}$ at the frequency of scattered light $\omega'$ and at that of the incident light $\omega$. 
It is the interference of these two contributions, which can be both constructive and destructive for objects with resonant permittivity, that leads to the strong directivity of the scattered light.

%%%%%%%%%%%%%%%%%%%%%%%%%%%%%%%%%%%%%%%%%%%%%
\begin{figure}[tb]
  \includegraphics[width=1.\columnwidth]{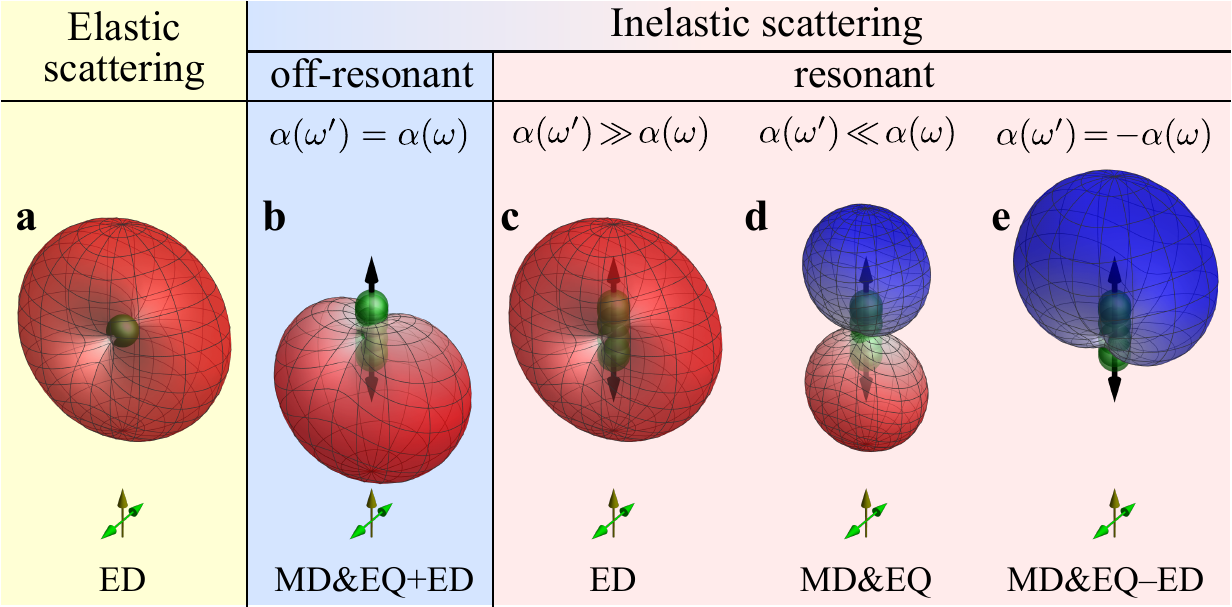}
\caption{{\bf Radiation pattern for  elastic and inelastic light scattering by a trembling particle}. 
Panel (a) shows the elastic scattering pattern, panels (b-e) describe inelastic scattering for
different ratios of  the polarizabilities at initial and scattered frequencies $\alpha(\omega)$ and $\alpha(\omega')$. The interference of the electric dipole (c) and magnetic dipole and quadrupole (d) patterns results in directional inelastic forward (e) and backward (b) scattering.
The light  is incident from the bottom (yellow arrow) and is linearly polarized (green arrow), the particle trembles along the light propagation direction (black arrow). Red and blue colors indicate the sign of electric field. 
}\label{fig:dia}
\end{figure}
%%%%%%%%%%%%%%%%%%%%%%%%%%%%%%%%%%%%%%%%%%%%%%

For the electric field at the anti-Stokes-shifted frequency $\omega'=\omega+\Omega$ at $r \to \infty$ we get
\begin{align}\label{eq:E}
\bm E'(\bm r) &=   \frac{\rmi\omega'^2 \e^{\rmi \omega' r /c}}{c^3r}  \big[ 
\omega' (\bm n_0 \cdot \bm u) \bm\alpha(\omega')\bm E_0 - \omega (\bm n \cdot \bm u) \bm\alpha(\omega) \bm E_0 \nonumber\\
&  -\Omega (\bm u \cdot  \bm E_0) \bm\alpha(\omega') \bm n_0  -\Omega ( \bm n  \cdot \bm\alpha(\omega) \bm E_0) \bm u 
\big]_\perp,
\end{align}
where $\bm n = \bm r/r$, $\bm n_0$ and $\bm E_0$ are the  propagation direction and electric field of the incident wave, subscript $\perp$ indicates that the perpendicular component with respect to $\bm n$ should be taken, $[\bm E]_\perp = -\bm n \times (\bm n \times \bm E)$.  
The field Eq.~\eqref{eq:E} can be decomposed into electric dipole $\bm p$, quadrupole $\bm Q$, and magnetic dipole $\bm m$ contributions oscillating at the frequency $\omega'$ with the amplitudes 
\begin{align}
&\bm d = \tfrac{\rmi}c \bm\alpha(\omega') \left[ \omega (\bm n_0 \cdot \bm u) \bm E_0  -  \Omega \bm u \times (\bm n_0 \times \bm E_0)  \right] \,, \nonumber\\
& \bm Q = 3 \bm\alpha(\omega)\bm E_0 \otimes \bm u + 3 \bm u  \otimes  \bm\alpha(\omega)\bm E_0  - 2\bm I (\bm u \cdot \bm\alpha(\omega)\bm E_0) ,  \nonumber \\
&\bm m= \tfrac{\rmi}{2c} \left(\omega- \Omega \right) [\bm\alpha(\omega)\bm E_0 \times \bm u]   \label{eq:muls}\,,
\end{align} 
where $(\bm a \otimes \bm b)_{\alpha\beta} = a_\alpha b_\beta$ and $\bm I$ is the identity matrix. 
All the multipole terms are of the same order, however, the induced electric dipole is proportional to the polarizability at the scattered light frequency $\bm\alpha(\omega')$ while electric quadrupole and magnetic dipole are determined by $\bm\alpha(\omega)$.
 Therefore, the frequency dependence of polarizability can be exploited to tune $\bm d$, $\bm m$, and $\bm Q$ to the Kerker condition.

%%%%%%%%%%%%%%%%%%%%%%%%%%%%%%%%%%%%
\begin{figure*}[tb]
  \includegraphics[width=.95\textwidth]{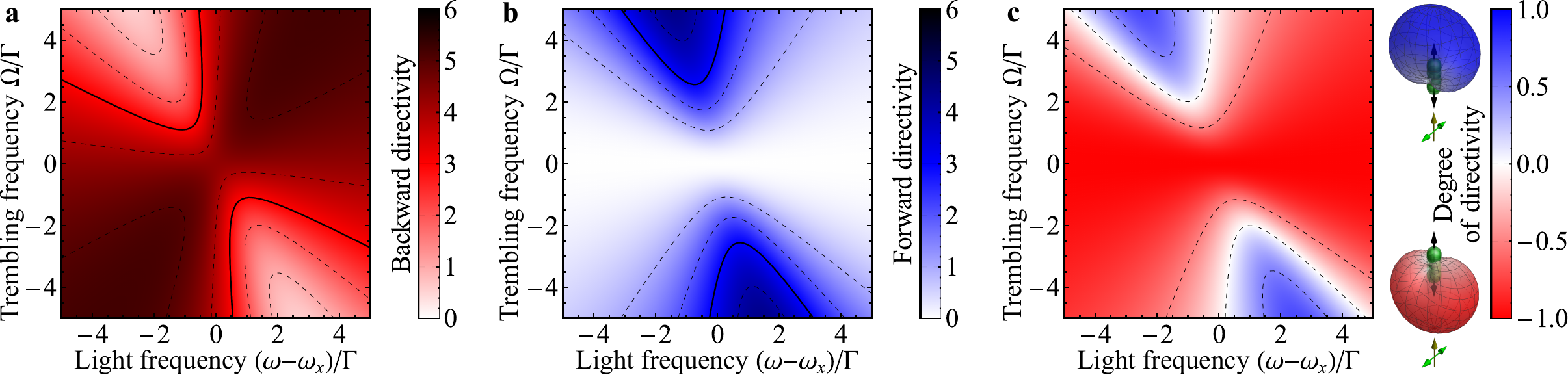}
\caption{{\bf Directivity of light scattered by a trembling resonant particle}. Panels 
{\bf (a)}  and {\bf (b)} show backward and forward directivity of non-polarized light depending  on the incident light frequency $\omega$ and trembling frequency $\Omega$. Solid line shows the directivity equal to 3 that limits usual Kerker effect. {\bf (c)} Degree of directivity $[D(\bm n_0)-D(-\bm n_0)]/[D(\bm n_0)+D(-\bm n_0)]$.
}\label{fig:freq}
\end{figure*}
%%%%%%%%%%%%%%%%%%%%%%%%%%%%%%%%%%%%%

\subsection{Optomechanical Kerker effect} 

Now we analyze in detail the direction pattern Eq.~\eqref{eq:E}  for light scattered on trembling particle with the isotropic resonant electric dipole polarizability $\alpha(\omega)$. We focus on the anti-Stokes component at the frequency $\omega'=\omega+\Omega$. Similar results for the Stokes component are obtained by inverting the sign of $\Omega$ and complex conjugation of the particle displacement vector $\bm u$. First, we neglect the last two terms in the right-hand-side of Eq.~\eqref{eq:E} proportional to the parameter $\Omega/\omega$, that is small for realistic systems. Figure~\ref{fig:dia} shows the radiation pattern of the  light scattered on the particle trembling along the propagation direction of the linearly polarized incident wave. Panel (a) shows the usual elastic electric dipole scattering at the frequency $\omega$, while  panels (c-d) correspond to the inelastic scattering to the frequency $\omega'$.
Panel (c) shows the contribution of the first electric dipole term in Eq.~\eqref{eq:E} to the scattered field, while panel (d) corresponds to the second term in Eq.~\eqref{eq:E} and a combination of magnetic dipole and electric quadrupole radiation. The total scattering intensity is a superposition of the patterns Fig.~\ref{fig:dia}(c) and Fig.~\ref{fig:dia}(d) with the coefficients $\alpha(\omega')$ and $\alpha(\omega)$, respectively. While the frequencies $\omega$ and $\omega'$ are close, the corresponding polarizabilities can differ strongly in the vicinity of the material resonance. Figures~\ref{fig:dia}(b) and~(e) show the two limiting cases when $\alpha(\omega')=\pm \alpha(\omega)$. In the non-resonant case,  $\alpha(\omega')= \alpha(\omega)$, the interference of electric dipole, magnetic dipole, and electric quadrupole radiation leads to the complete suppression of forward inelastic scattering, Fig~\ref{fig:dia}(c). In the opposite case of strong frequency dependence of polarizability when  $\alpha(\omega')= -\alpha(\omega)$, Fig.~\ref{fig:dia}(d), the inelastic backscattering vanishes.

Using Eq.~\eqref{eq:E} we evaluate the cross-section of the inelastic scattering  for unpolarized light
\begin{align}
\frac{d\sigma}{do} = \frac{\omega^6}{2c^6} \big|[\alpha(\omega')\bm n_0 - \alpha(\omega) \bm n]\cdot \bm u\big|^2 [1+(\bm n_0 \cdot \bm n)^2] \,,\label{eq:sigma}
\end{align}
where $do$ is the solid angle for scattered light direction. In the non-resonant case when 
$\alpha(\omega') = \alpha(\omega)$ and $\bm n_{0}\parallel \bm u$ we recover the geometric optics result  Eq.~\eqref{eq:intro}  with $d\sigma \propto (1-\cos\theta)^{2}$ and suppressed forward scattering.

 The asymmetry of the light scattering pattern can be quantified by the directivity $D(\bm n) = 4 \pi (\int d\sigma)^{-1} d\sigma/do  $ ~\cite{Krasnok2013}. In the considered geometry, $\bm u \parallel \bm n_0$, the directivity  for forward ($\bm n = \bm n_0$) and backward ($\bm n = -\bm n_0$) scattering reads
\begin{align}
D(\pm\bm n_0) = \frac{15 |\alpha(\omega') \mp \alpha(\omega)|^2}{10|\alpha(\omega')|^2 +4 |\alpha(\omega)|^2}\:.
\end{align}
For the non-resonant case when $\alpha(\omega')= \alpha(\omega)$ the forward scattering is absent while the backward directivity reaches $30/7$. The maximal value of forward (backward) directivity is $5.25$  that is achieved when $\alpha(\omega') =\pm (2/5)\, \alpha(\omega)$. Thus, the directivity of the optomechanical Kerker effect surpasses the limiting value of 3 for the classical Kerker effect, because the electric quadrupole contribution is additionally involved~\cite{Liu2018}.

For numerical demonstration we consider the simplistic general model of the  particle characterized by the resonant polarizability 
\begin{align}
\alpha(\omega) = \frac{A}{\omega-\omega_x + \rmi \Gamma} \,,
\end{align}
where $A$ is a constant, $\omega_x$ is the resonance frequency, and $\Gamma$ is the resonance width. Such dependence corresponds to, e.g., exciton resonance in quantum dots and transitional metal dichalcogenide monolayers, electron transitions in cold atomic gases, plasmon resonance in graphene, resonances in atomic nuclei probed by M\"ossbauer $\gamma$-ray spectroscopy, superconducting resonators for radio-frequency electromagnetic field, see Discussion. 
Figures~\ref{fig:freq}(a) and~\ref{fig:freq}(b) show the color plots of the directivity of backward and forward anti-Stokes light scattering depending on the incident light frequency $\omega$ and the trembling frequency $\Omega$. In the dark areas bounded by the solid lines the directivity is larger than 3, which can be termed as a  super-Kerker effect. For backward and forward scattering it is realized  when initial and scattered light frequencies $\omega$ and $\omega'=\omega+\Omega$ are  located on the same side or on the opposite sides of the resonance, respectively. Shown in the Fig.~\ref{fig:freq}(c) is the degree of directivity $[D(\bm n_0)-D(-\bm n_0)]/[D(\bm n_0)+D(-\bm n_0)]$. Red and blue colors indicate predominance of the backward and forward scattering, respectively. Degree of directivity reaches $\pm 1$ if $\alpha(\omega') = \mp \alpha(\omega)$, that is realized at $\Omega = 2(\omega_x-\omega) \gg \Gamma$ and $\Omega \to 0$, respectively.

%%%%%%%%%%%%%%%%%%%%%%%%%%%%%%%%%%%%%
\begin{figure}[t]
  \includegraphics[width=.99\columnwidth]{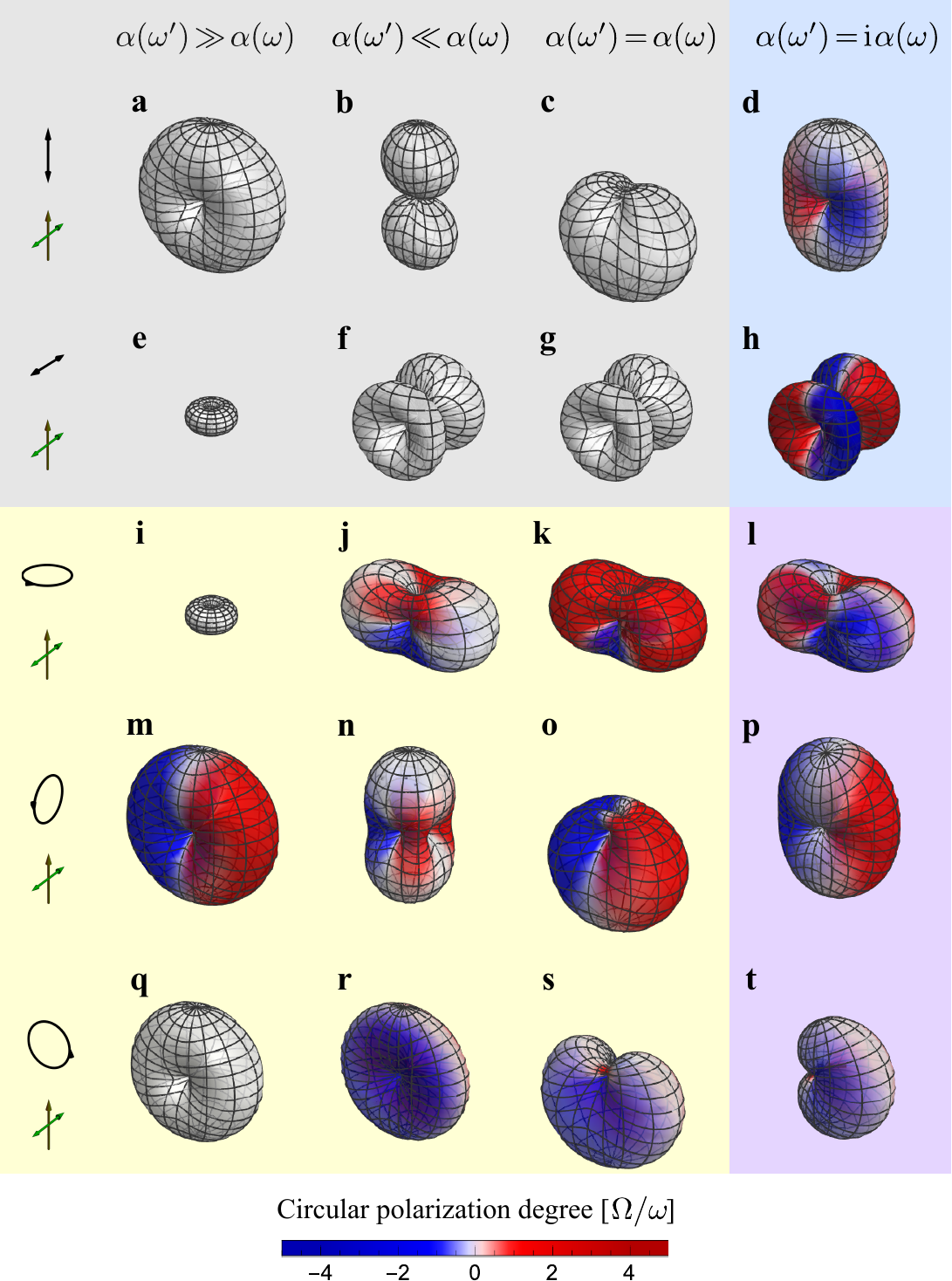}
\caption{{\bf Optomechanical Spin Hall effect}. The anti-Stokes light scattering pattern with the degree of circular polarization marked by red and blue colors. Light is incident from the bottom and linearly polarized, see yellow and green arrows on the left, respectively.
Rows correspond to different trembling directions indicated by black arrows on the left. Columns correspond to different relations between particle polarizabilities at the frequencies of incident and scattered light, $\alpha(\omega)$ and $\alpha(\omega')$, indicated in the top. 
The colored areas of the table represent different origin of the optomechanical spin Hall effect: phase difference between the polarizabilities $\alpha(\omega)$ and $\alpha(\omega')$ (blue color), the phase difference between components of the displacement vector $\bm u$ (yellow color), or their combined action (purple color). 
}\label{fig:circ}
\end{figure}
%%%%%%%%%%%%%%%%%%%%%%%%%%%%%%%%%%%%%

\subsection{Optomechanical Spin Hall effect}
The interference of the electric and magnetic modes is known to give
rise  to a strong circular polarization of the scattered light upon excitation with a linearly polarized light~\cite{Glazov2005,Leyder2007}. Conversely, photons with opposite circular polarizations scatter in different directions. This is termed as an optical spin Hall effect in analogy with the spin-dependent scattering of electrons in solids~\cite{Dyakonov2008}. Here, we put forward an {\it optomechanical spin Hall effect}, i.e.,  inelastic polarization-dependent directional scattering on a trembling particle.

The two first  terms in Eq.~\eqref{eq:E} are dominant  and yield the scattered light with the same polarization as the incident. Optomechanical spin Hall effect results from  the last two terms, that give a small correction of the order $\Omega/\omega$ describing linear-to-circular polarization conversion.
 The circular polarization degree of the plane wave with the electric field amplitude $\bm E$ can be defined as $P_c(\bm n) = \rmi \bm n \cdot [\bm E \times \bm E^*]/ |\bm E|^2$. Substituting here  the scattered wave from Eq.~\eqref{eq:E} we obtain for the case of non-polarized incident light
\begin{align}\label{eq:Pc}
P_c = \frac{\Omega}{\omega} \, \frac{2\,\bm n \times \bm n_0}{1+(\bm n_0 \cdot \bm n)^2} \cdot \text{Im\,}\frac{[\alpha(\omega')+(\bm n_0 \cdot \bm n)\alpha(\omega)]\bm u}{[\alpha(\omega') \bm n_0 - \alpha(\omega)\bm n]\cdot \bm u }\:,
\end{align} 
where we keep linear in $\Omega/\omega$ terms only. Equation~\eqref{eq:Pc} indicates two possible origins of circular polarization: (i) the phase difference of the polarizabilities $\alpha(\omega)$ and $\alpha(\omega')$ and (ii) the phase difference of the components of the displacement vector $\bm u$. The first mechanism is likely to contribute in the vicinity of the material resonance where the phase of $\alpha$ rapidly changes by $\pi$. The second mechanism is realized even away from the  resonances, however it requires the particle trembling around a circle or an ellipse rather  than  just along one axis.

Figure~\ref{fig:circ} shows by red and blue color the circular polarization degree of the anti-Stokes-scattered light for different particle trembling directions, indicated on the left, and different relations between polarizabilities at the frequencies of incident and scattered light, indicated on the top. The plots on the gray background  show the cases where both mechanisms (i) and (ii) are absent, so the circular polarization does not emerge.  
The first mechanism is realized for the plots on the blue background, where we as an example assume $\alpha(\omega') = \rmi \alpha(\omega')$. For $\bm u \parallel \bm n_0$, see Fig.~\ref{fig:circ}(d), the dependence of $P_c$ on  the azimuthal  angle is described by the second angular harmonic, so $P_c$ is inverted when the incident light with the perpendicular polarization is considered. For unpolarized excitation, the circular polarization vanishes in agreement with Eq.~\eqref{eq:Pc}. 
Figure~\ref{fig:circ}(h) shows the angular pattern of  $P_c$ for light scattered by the particle trembling perpendicularly to the direction of incident light and  $\bm E_0 \parallel \bm u$.  For the other linear polarization of the incident light (not shown) the conversion to circular polarization is absent. Therefore, even for the non-polarized incident light the circular polarization of scattered light persists and it is described by Eq.~\eqref{eq:Pc}. 

Now we turn to the second mechanism  of the generation of circular polarization, that is realized for the plots on the yellow background. Figures~\ref{fig:circ}(i)-(k) illustrate the  circular polarization of the light scattered by the particle trembling around a circular trajectory in the plane perpendicular to the incident light direction. 
Then, the forward and backward scattered light reveal opposite signs of circular polarization, except for the case of Fig.~\ref{fig:circ}(i) when the scattered wave is of the order of the small parameter $\Omega/\omega$ and linearly polarized, see Eq.~\eqref{eq:E}. Figures~\ref{fig:circ}(m)-(o) and~(q)-(s) show the pattern of the circular polarization degree of the light scattered by the particle trembling around a circular trajectory in the plane parallel to the incident incident light propagation direction. The  circular polarization sign depends on  whether  the light is scattered to the left or to the right with respect to the plane of trembling.
Finally, if both optomechanical Spin Hall effect mechanisms  are present, see plots on the purple backgound, their interplay leads to a strong asymmetry of both the scattering intensity pattern and the circular polarization pattern.

%%%%%%%%%%%%%%%%%%%%%%%%%%%%%%%%%%%%%%%%%%%%%%%%%%%

%%%%%%%%%%%%%%%%%%%%%%%%%%%%%%%%%%%%%
\begin{figure*}[tb!]
\includegraphics[width=.9\textwidth]{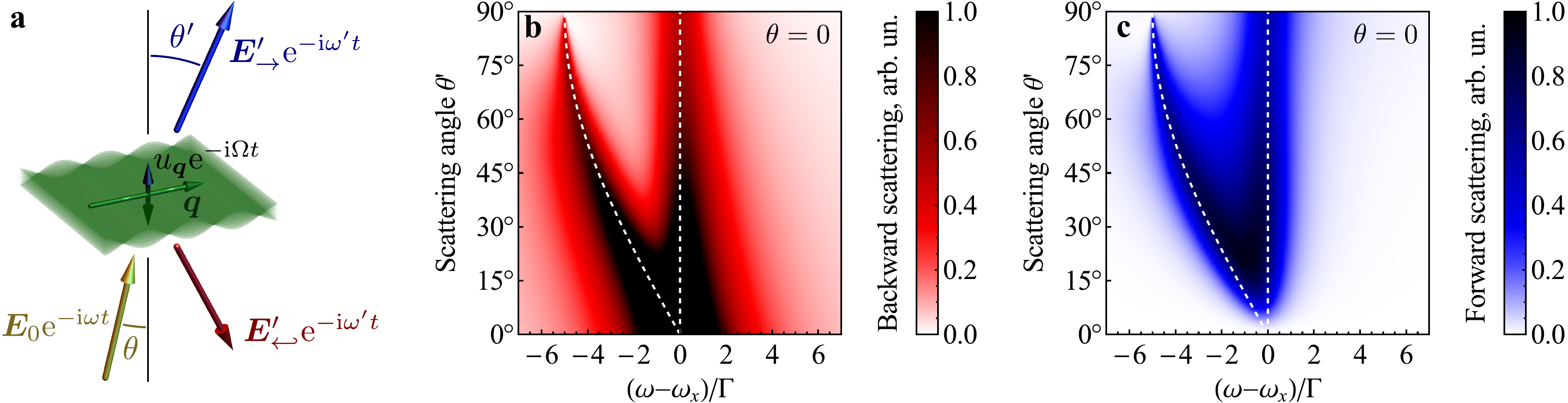}
\caption{{\bf Optomechanical Huygens surface.} (a) Light scattering on a resonant layer with flexural vibrations. Color maps of (b) backward and (c) forward scattered power for the case of normal incidence as a function of incident light frequency and scattering angle $\theta'$. Dashed lines indicate the resonances for incident and scattered light. The reflection coefficient of resonant layer was taken in the form $r_{s(p)}(\omega,\theta) = -\rmi\Gamma_{s(p)}/(\omega-\omega_x+\rmi\Gamma_{s(p)})$, with $\Gamma_s = \Gamma/\cos \theta$ and $\Gamma_p = \Gamma \cos \theta$~\cite{Ivchenko2005}.
The linear dispersion of phonons was assumed, $\Omega = s|\bm q|$, with $(s/c)(\omega_x/\Gamma) = 5$.
}\label{fig:huyg}
\end{figure*}
%%%%%%%%%%%%%%%%%%%%%%%%%%%%%%%%%%%%%

\subsection*{Optomechanical Huygens surfaces}
A two-dimensional planar array of scatterers tuned to the Kerker condition is known to demonstrate no forward or backward scattering~\cite{Staude2013,Pfeiffer2013,Ding2018}. When the backscattering is suppressed it realizes a Huygens' surface that transmits light without reflection and modifies only it phase. An optomechanical analog of the Huygens' surface is a thin layer with a resonant dipole polarizability that trembles along its normal, see Fig.~\ref{fig:huyg}(a). 

Since the flexural layer vibrations can possess an in-plane wave vector $\bm q$, the direction of scattered light can differ from that of the incident light. It is determined from the conservation of the in-plane wave vector component in the process of anti-Stokes (Stokes) scattering, $\bm k' = \bm k \pm \bm q$, where $k=(\omega/c) \cos \theta$ and $k'=(\omega'/c) \cos \theta'$ are the in-plane wave vectors of the incident and scattered light, $\theta$ and $\theta'$ are the angles between the light propagation direction and the layer normal. 
The electric field amplitudes of the anti-Stokes forward ($\rightarrow$) and backward ($\hookleftarrow$) scattered light are given by the diagrams of Fig.~\ref{fig:dtech}(b)-(c) and read $E'^{\rightarrow(\hookleftarrow)}_{\sigma} = \sum_{\sigma'}S^{\rightarrow(\hookleftarrow)}_{\sigma\sigma'} E_{0,\sigma'} u_{\bm q,z} $, where $\sigma,\sigma'$ enumerates two polarizations, $s$ and $p$. The Jones matrix $S^{\rightarrow(\hookleftarrow)}_{\sigma\sigma'}$ with the elements (see Methods for calculation details)
\begin{align}\label{eq:huyg}
&S^{\rightarrow(\hookleftarrow)}_{ss} = \rmi \frac{\omega'}{c}\, \cos\theta \cos\phi \left[ r_s(\theta',\omega') \mp r_s(\theta,\omega)  \right] ,\\\nonumber
&S^{\rightarrow(\hookleftarrow)}_{ps} = \rmi \frac{\omega'}{c}\, \frac{\cos\theta \sin\phi}{\cos\theta'} \left[ r_p(\theta',\omega') \mp r_s(\theta,\omega)  \right] ,\\\nonumber
&S^{\rightarrow(\hookleftarrow)}_{sp} = -\rmi \frac{\omega'}{c}\, \sin\phi \left[ r_s(\theta',\omega') \mp r_p(\theta,\omega)  \right] ,\\\nonumber
&S^{\rightarrow(\hookleftarrow)}_{pp} = \rmi \frac{\omega'}{c}\, \frac{\cos\phi -\sin\theta \sin\theta'}{\cos\theta'} \left[ r_p(\theta',\omega') \mp r_p(\theta,\omega) \right] ,
\end{align}
describes polarization conversion, $r_s(\theta,\omega)$ and $r_p(\theta,\omega)$ are the reflection coefficients  for oblique incidence of $s$- and $p$-polarized light on the layer at rest, $\phi$ is the angle between the in-plane wave vectors $\bm k$ and $\bm k'$. Similarly to the optomechanical Kerker effect for the trembling particle, the forward (backward) scattering on a trembling layer vanishes when $r_{s(p)}(\theta',\omega')  = \pm r_{s(p)}(\theta,\omega)$. The power of the anti-Stokes light  scattered forward (backward) into the solid angle $do$ by the unit area of the layer for the case of unpolarized excitation with the intensity $I_0$  reads
\begin{align}
\frac{dI'}{do} =     \frac{\omega'^2}{c^2}  \frac{\cos^2 \theta'}{\cos \theta}\, \frac12 \sum_{\sigma\sigma'}|S^{\rightarrow(\hookleftarrow)}_{\sigma\sigma'}|^2\; |u_{\bm q}|^2 I_0\:.
\end{align}
Figures~\ref{fig:huyg}(b)-\ref{fig:huyg}(c) show the forward and backward scattered power for the case of normal incidence, $\theta=0$, and constant $|u_{\bm q}|^2$. Both plots feature two resonances indicated by dashed lines: the resonance for incident light at $\omega=\omega_x$ and the resonance for scattered light at $\omega+\Omega = \omega_x$. Since the vibration frequency $\Omega$ increases with the vibration wave vector $q = (\omega'/c) \sin \theta'$, the two resonances split with an increase of the scattering angle $\theta'$. The main result of Figs.~\ref{fig:huyg}(b)-(c) is that the forward scattering involving vibration with $\bm q =0$, i.e., for $\theta'=0$, is suppressed while the backward scattering at $\theta'=0$ is increased. The absence of the forward-scattered wave in the limit $\Omega, \bm q \to 0$ has a clear physical interpretation: in the quasi-static picture, the shift of the layer as a whole affects the reflected plane wave but does not affect the transmitted one.

%%%%%%%%%%%%%%%%%%%%%%%%%%%%%%%%%%%%%%%%%%%%
\section{Discussion}

%%%%%%%%%%%%%%%%%%%%%%%%%%%%%%%%%%%%%%%%%%%%

\begin{table*}[t]
\begin{center}
\begin{tabular}{|l||c|r|r|r||c|c|c|}
\hline
System   & $\hbar\omega_x$ & $\Gamma/(2\pi)$\hphantom{[~??]} & $\Omega/(2\pi)$\hphantom{??]} & $u$\hphantom{[????]} & Tunability $\frac{\Omega}{\Gamma}$ & Coupling $\frac{\omega_x}{c}u$ \\
\hline \hline 
Plasmon in graphene~\cite{Yan2013,Chen2009,Davidovikj2017} & 0.1\,--\,1\,eV & 10\,THz & 0.1\,GHz & 10\,nm & $10^{-5}$ & 0.01\,--\,0.1  \\\hline
Exciton in TMD monolayers~\cite{Samanta2015,Davidovikj2017} & 2\,eV & 20\,GHz & 0.1\,GHz &10\,nm & 5$\cdot$$10^{-3}$ & 0.1  \\\hline
Exciton in colloidal QDs~\cite{Ferne2014,Saffar2014} &  2\,eV & 400\,MHz & 20\,MHz & 200\,nm & 5$\cdot$$10^{-2}$ & 2  \\\hline
Cold atoms~\cite{Jessen1992} & 1.5\,eV & 10\,kHz & 100\,kHz & 200\,nm & 10 & 1.5 \\\hline
Superconducting qubits~\cite{Paik2011,Saffar2014} & 20\,$\mu$eV & 0.1\,MHz & 20\,MHz  & 200\,nm & 200 & 2$\cdot$$10^{-5}$  \\\hline
Resonance in nuclei~\cite{Smirnov1977,Dewar1987} & 15\,keV & 0.5\,MHz & 10\,GHz & $10^{-3}$\,nm & 2$\cdot$$10^4$& 0.1 \\\hline
\end{tabular}
\end{center}
\caption{Parameters of various resonant optomechanical systems}\label{table:systems}
\end{table*}

%%%%%%%%%%%%%%%%%%%%%%%%%%%%%%%%%%%%%

%%%%%%%%%%%%%%%%%%%%%%%%%%%%%%%%%%%%%%%%%
\begin{figure}[b]
\includegraphics[width=.45\textwidth]{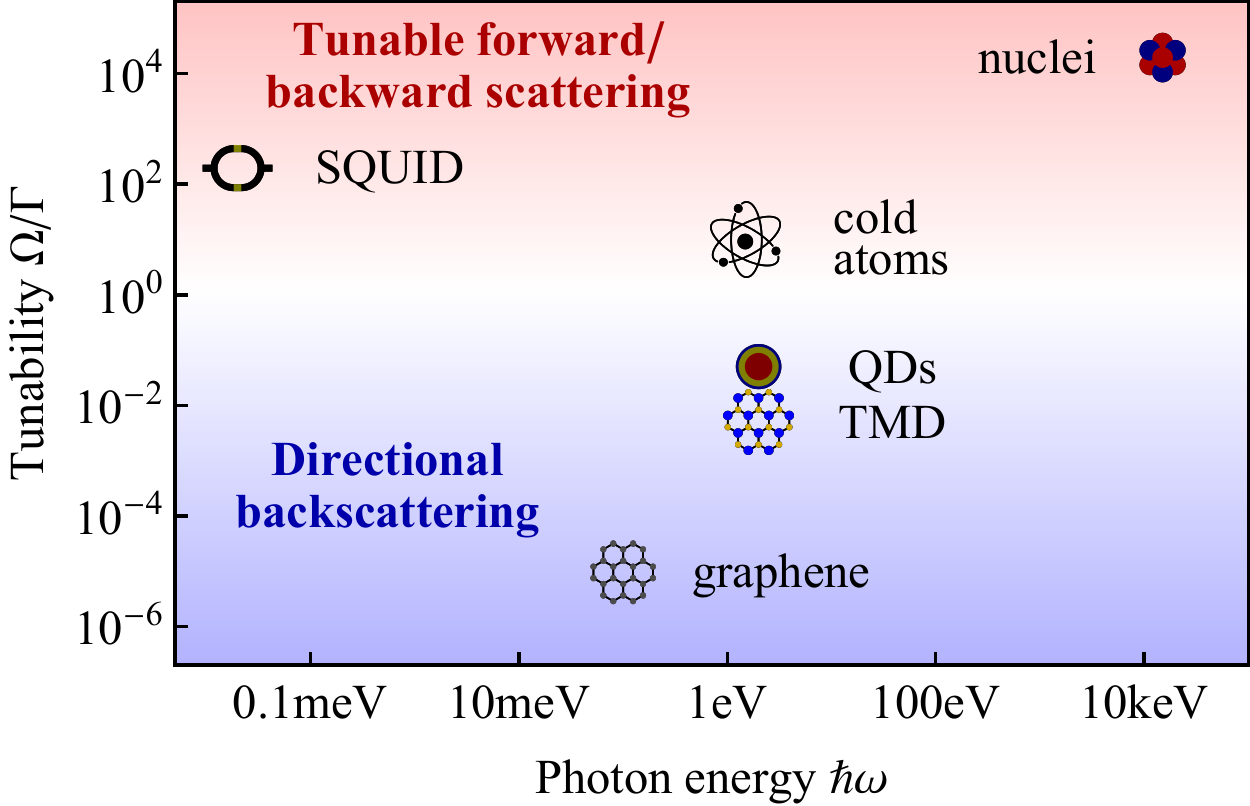}
\caption{{\bf Resonant optomechanical systems}. Shown is  the tunability of resonant scatterers operating in different spectral ranges.  When $\Omega \ll \Gamma$ (blue area), the directional scattering can occur only in the backward direction.  When $\Omega \gg \Gamma$ (red area), one can realize both directional forward and backward scattering by a proper tuning of the incident light frequency and trembling frequency.  Numerical parameters and references are given in Table~\ref{table:systems}.}\label{fig:params}
\end{figure}
%%%%%%%%%%%%%%%%%%%%%%%%%%%%%%%%%%%%%

Now we discuss the possibilities for  experimental realization of the optomechanical Kerker effect.
The proof-of-principle observation of the suppression of the forward inelastic scattering in the nonresonant regime, Fig.~\ref{fig:dia}(b), seems to be relatively straightforward for an arbitrary subwavelength particle.  The only requirements are to operate in the far field regime and to avoid the internal deformations  of the particle, so that it trembles as a whole.  The true challenge is to realize a dynamical tunability between forward and backward scattering by exploiting the resonance of particle polarizability. This requires the width of the resonance $\Gamma$ to be comparable with the frequency of vibrations $\Omega$, i.e., narrow resonances and high vibration frequencies. Table~\ref{table:systems}  presents an overview of various potential systems in different ranges of electromagnetic spectrum and Fig.~\ref{fig:params} visualizes  their tunability $\Omega/\Gamma$.
Apparently, the highest degree of tunability can be attained by exploiting the M\"ossbauer resonances in the nuclei of crystals for  $\gamma$-rays~\cite{rohlsberger1999b}. Namely, the linewidth can be as narrow as 0.5~MHz~\cite{Smirnov1977}, while the GHz range hypersound waves in metals are available~\cite{Dewar1987}. Further opportunities are provided by organizing  planar cavities for synchrotron $\gamma$-radiation~\cite{Roehlsberger2010}. The opposite side of the electromagnetic spectrum, with sub 0.1\,meV rather than 10\,keV photon energies, is represented by the superconducting qubits~\cite{Paik2011}. They feature high quality factor resulting in potentially high tunability when being coupled to the ultrasound generator \cite{Saffar2014}. An apparent drawback of such setup is a relatively weak scattering efficiency due to the  vibration amplitude being relatively small as compared to the electromagnetic wavelength, see Table~\ref{table:systems}.
The THz and optical frequency ranges are accessible by membranes made of graphene~\cite{Chen2009}  and transition metal dichalcogenide monolayers~\cite{Samanta2015}, respectively. These platforms feature reasonable coupling strength but have limited tunability because of the relatively low frequency of the flexural vibrations ($\lesssim 0.1~\rm GHz$) as compared to the broad width of plasmonic or excitonic resonance.
High tunability and strong optomechanical coupling efficiency for visible light can be realized by exploiting narrow resonances in cold atoms
vibrating in an optical trap \cite{Jessen1992}. Alternatively, one could use excitonic resonances in artificial atoms, colloidal quantum dots \cite{Saffar2014}.

The rich consequences of the interplay of magnetic and electric response on the electromagnetic wave propagation are known at least for 50 years since the seminal work by Veselago on the media with negative permittivity and permeability~\cite{Veselago1967,veselago2006}.
Still, the mutual effect of electric and magnetic resonances is  very far from being completely understood. For example, it has been realized only quite recently that the interference  and coupling of 
 electric and magnetic resonances underpin bianisotropic photonic topological insulators \cite{Khanikaev2013,Slobozhanyuk2016}, where the light backscattering on disorder is suppressed.
We expect that the proposed   optomechanical Kerker and spin-Hall effects with trembling-induced magnetic response open  a pathway to engineer chiral optomechanical coupling at nanoscale, expanding the chiral quantum optics \cite{Lodahl2017,Spitzer2018} to the optomechanical domain. Our results  can be instructive for the design of nonreciprocal topological  circuits \cite{Lu2016,Ozawa2018}, where the disorder-robust  propagation of light and sound is ensured by the time modulation of optical and mechanical properties~\cite{Sounas2017,Poshakinskiy2017,Sohn2018}.

%\newpage
%\clearpage
\appendix

%%%%%%%%%%%%%%%%%%%%%%%%%%%%%%%%%%%%%%%%%
\setcounter{equation}{0}
\renewcommand{\theequation}{M\arabic{equation}}
\setcounter{figure}{0}
\renewcommand{\thefigure}{M\arabic{figure}}
\section{Methods}
\small 

\subsection{Polarization of trembling media}

We consider the light with the frequency $\omega$ incident on a medium vibrating with the frequency $\Omega$. Medium motion is described by the displacement vector $\bm u(\bm r, t) = \bm u (\bm r) e^{-\rmi \Omega t} + \text{c.c.}$ 
We suppose that the vibration amplitude is small and focus on the linear-in-$\bm u$ effect only, i.e. appearance of  polarization at anti-Stokes- and Stokes-shifted frequencies $\omega \pm \Omega$. 

Consider the unitary volume of medium that in the absence of vibration had coordinate $\bm r$. Its polarization at the time $t$ is determined by the electric field $\widetilde{\bm E}(\bm r , t')$ that has acted on it in all previous moments of time $t' < t$,
\begin{align}\label{eq:tP}
\widetilde {\bm P}(\bm r,t) = \int_{-\infty}^t dt' \, \bm\chi(\bm r,t-t') \widetilde{\bm E}(\bm r , t'),
\end{align}
where $\bm\chi (\bm r, \tau)$ is the dielectric susceptibility function. We do not account here for the possible change of susceptibility under medium deformation, because such photoelastic effect requires separate microscopic calculation. While photoelasticity may give dominant contribution to optomechanical coupling in resonant structures~\cite{Berstermann2009,Fainstein2013,Jusserand2015,Poshakinskiy2016PRL}, it does not play any role in the effects we consider, where the objects move as a whole and deformation is absent. 
The electric field $\widetilde{\bm E}(\bm r , t')$ in Eq.~\eqref{eq:tP} should be calculated in the reference frame that moves and rotates together with the considered medium volume. Keeping  linear-in-$\bm u$ terms only we obtain
\begin{align}\label{eq:tE}
\widetilde{\bm E} =\bm E + \left( \bm u \frac{\partial}{\partial \bm r}\right) \bm E - \frac{\text{rot\,}\bm u}2 \times \bm E +\frac1c \frac{\partial \bm u}{\partial t} \times \bm B \,,
\end{align}
where $\bm E$ and $\bm B$ are the electric and magnetic fields in the reference frame at rest, and all quantities are evaluated at the moment $t'$. Second term in the right-hand-side of Eq.~\eqref{eq:tE} stems from the fact that the electric field should be evaluated at the point $\bm r + \bm u(\bm r, t')$, third term accounts for the medium rotation% 
, and the last term comes from the Lorentz transform.

Equation~\eqref{eq:tP} gives the polarization of the unitary volume of undeformed medium in the reference frame that moves and rotates with the medium. In the reference fame at rest, the polarization density reads
\begin{align}\label{eq:P}
\bm P=\widetilde{\bm P} -  \left( \bm u \frac{\partial}{\partial \bm r}\right) \widetilde{\bm P} +\frac{\text{rot\,}\bm u}2 \times \widetilde{\bm P}- \widetilde{\bm P} \, \text{div\,} \bm u \,,
\end{align}
where the last term accounts for the difference between the   deformed and undeformed unitary volumes.
Additionally, the magnetization $
\bm M =  - {\partial \bm u}/{\partial t} \times \widetilde{\bm P} $
appears in the frame at rest due to the Lorentz transform.

Finally, we combine Eqs.~\eqref{eq:tP}--\eqref{eq:P}  and evaluate the current $\bm j = \partial \bm P/{\partial t} + \text{rot\,} \bm M $ in the reference frame at rest. The relation between the current $\bm j$ at the anti-Stokes-shifted frequency $\omega' = \omega+\Omega$ and the vector potential $\bm A$ of light at the initial frequency $\omega$ in the $\bm k$-space assumes the form 
$\bm j_{\bm k'}(\omega') = \sum_{\bm k} \delta\bm\Pi_{\bm k', \bm k} \bm A_{\bm k}(\omega)$,
where we use the gauge with zero scalar potential,
\begin{align}\label{eq:dP}
\delta\bm\Pi_{\bm k', \bm k} &= \omega^{\prime 2} \bm\chi_{\bm k'+\bm q-\bm k}(\omega') \,\bm\Lambda_{\bm k+\bm q,\bm k} (\omega',\omega) \nonumber \\
 &+ \bm\Lambda^{\rm T}_{\bm q-\bm k',-\bm k'} (-\omega,-\omega') \, \omega^{2} \bm\chi_{\bm k'+\bm q-\bm k}(\omega) \,,
\end{align}
$\chi_{\bm q} (\omega) = \iint \chi(\bm r, \tau) \e^{\rmi \omega \tau -\rmi \bm q \cdot \bm r} d\tau\, d^3r$ and $\bm u_{\bm q}$ are the Fourier transforms of $[\varepsilon(\bm r,\omega)-1]/(4\pi)$ and  $\bm u(\bm r)$, respectively,  superscript $\rm T$ denotes matrix transposition, and 
\begin{align}\label{eq:L}
\bm\Lambda_{\bm k+\bm q,\bm k} (\omega',\omega) &=  \rmi (\bm u_{\bm q} \cdot \bm k)  - \rmi \frac{\omega'-\omega}{\omega'} \bm k \otimes \bm u_{\bm q} \nonumber\\
&-\rmi \frac{\omega}{\omega'} \frac{\bm u_{\bm q} \otimes \bm q - \bm q \otimes  \bm u_{\bm q}}2 \, .
\end{align}
To calculate the polarization at the Stokes-shifted frequency $\omega -\Omega$ one should change in the above equations the sign of $\Omega$ and replace $\bm u_{\bm q}$ with $\bm u_{-\bm q}^*$. 
The quantity $\delta\bm\Pi$ is the correction to the polarization operator caused by the medium vibration. The two terms in the right-hand side of Eq.~\eqref{eq:P} can be represented diagrammatically as shown in Figs.~\ref{fig:dtechS}(a) and~\ref{fig:dtechS}(b). The wavy line corresponds to a photon, dashed line is a vibration, the bubble stands for polarization operator of medium at rest, $\bm \Pi_{\bm k', \bm k}(\omega) = \omega^2 \bm\chi_{\bm k'-\bm k}(\omega)$, and solid dot represents optomechanical interaction $\bm\Lambda$.

\subsection{Light scattering on trembling objects}

\begin{figure}[t]
  \includegraphics[width=.99\columnwidth]{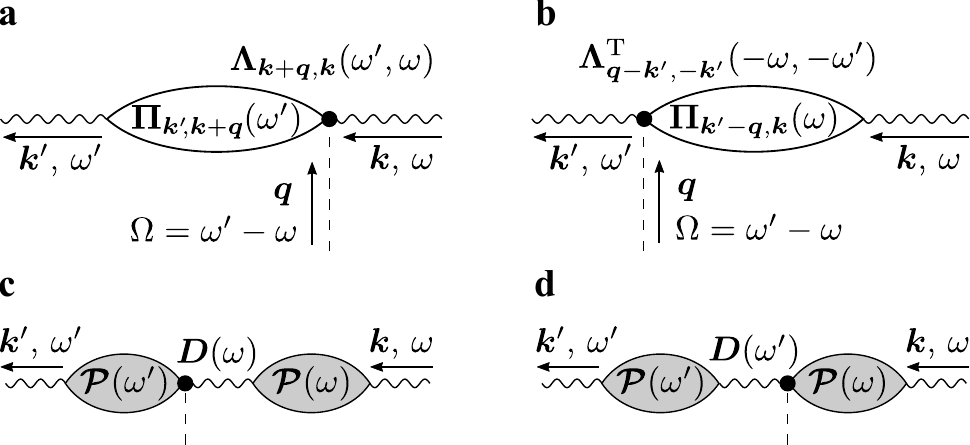}
\caption{{\bf Diagrammatic representation of the inelastic scattering.} (a) and (b) Diagrammatic representation for the optomechanical interaction of light (wavy line), medium polarization (bubble), and mechanical vibration (dashed line).  
(c) and (d) The contributions to the amplitude of light scattering by a trembling medium that together with contributions shown in Fig.~\ref{fig:dtech}(b)-(c) give the full amplitude of inelastic light scattering.
Wavy line denotes photon and corresponds to the Green's function for vector potential $\bm D(\bm k,\omega) = -4\pi (1 - \bm k \otimes \bm k/\omega^2) /(\omega^2- \bm k^2)$, empty and filled bubbles correspond to the bare and dressed polarization operators of the medium at rest, $\bm \Pi$ and $\bm{\mathcal{P}} = \bm \Pi (1-\bm D \bm \Pi)^{-1}$, respectively, dashed line represents mechanical displacement $\bm u_{\bm q}$, solid dot is the optomechanical interaction $\bm \Lambda$ given by Eq.~\eqref{eq:L}.}\label{fig:dtechS}
\end{figure}

Here we describe how the amplitude of inelastic light scattering on a trembling object of an arbitrary shape can be calculated. We use $c=1$ for simplicity. The full amplitude can be represented as a sum of four terms diagrammatically shown in Fig.~\ref{fig:dtech}(b)-(c) and Fig.~\ref{fig:dtechS}(c)-(d). They correspond  to the medium polarization, described by the dressed polarization operator $\bm{\mathcal{P}} = \bm \Pi (1-\bm D \bm \Pi)^{-1}$, accounted before, after, or both before and after the optomechanical interaction.

%%%%%%%%%%%%%%%%%%%%%%%%%%%%%%%%%%%%%%%%
%%%%%%%%%%%%%%%%%%%%%%%%%%%%%%%%%%%%%%%%

Note that photon Green's function $\bm D_{\bm k}(\omega) = -4\pi (1 - \bm k \otimes \bm k/\omega^2) /(\omega^2- \bm k^2)$ is $\bm k$-even while the optomechanical interaction $\bm \Lambda$ is  $\bm k$-odd in the absence of the last term describing medium rotation, see Eq.~\eqref{eq:L}. In the main text, we consider the small object characterized by the wave-vector-independent dressed polarization operator $\bm{\mathcal{P}}_{\bm k',\bm k}(\omega) = \omega^2 \bm\alpha(\omega)$. In such case, the summation over the phonon wave vectors in the intermediate states of the diagrams Fig.~\ref{fig:dtechS}(c) and~(d) yields zero. The scattering amplitude is then given by the remaining diagrams Fig.~\ref{fig:dtech}(b) and~(c). 

In the general case all the diagrams of Fig.~\ref{fig:dtechS} contribute to the total scattering matrix element that reads 
\begin{align}
M_{\bm k' \bm k} = 
\bm A_{\bm k'}^* \cdot  (1+\bm{\mathcal{P}D}) (\bm{\Lambda^T \Pi + \Pi\Lambda}) (1+\bm{D\mathcal{P}})\bm A_{\bm k} ,
\end{align}
where $\bm A_{\bm k} = \sqrt{2\pi/\omega} \,\bm e_0\, \e^{\rmi \bm k \cdot \bm r} $  and
$\bm A_{\bm k'} = \sqrt{2\pi/\omega'} \,\bm e'\, \e^{\rmi \bm k' \cdot \bm r}$
are the vector potentials of the incident and scattered photons, $\bm e_0$ and $\bm e'$ are their polarizations.
Introducing  the field distributions of incident and scattered photons in the form
$\bm{\mathcal{A}}_0 (\bm r)= (1+\bm{D\mathcal P}) \bm A_{\bm k}$ and
$\bm{\mathcal{A}}'^* (\bm r)= (1+\bm{D\mathcal{P}}^T) \bm A_{\bm k'}^*$
we finally obtain the scattering amplitude per solid angle $do$
\begin{align}\label{eq:S:S}
\frac{dS_{\bm k' \bm k}}{do} = \frac{\rmi \omega'^2}{(2\pi)^2} \int \big[ \bm\Pi^T(\bm r,\omega')\bm{\mathcal{A}}'^* (\bm r) \cdot \bm\Lambda (\omega',\omega) \bm{\mathcal{A}}_0 (\bm r) \nonumber\\
 + \bm\Lambda (-\omega,-\omega') \bm{\mathcal{A}}'^* (\bm r) \cdot  \bm\Pi(\bm r,\omega) \bm{\mathcal{A}}_0 (\bm r) \big] d\bm r
\,. 
\end{align}
Here, the operator $\bm \Lambda$ with the components
\begin{align}\label{eq:S:L}
\Lambda_{\alpha\beta} (\omega',\omega) =  \left( \bm u \cdot \frac{\partial}{\partial \bm r} \right) \delta_{\alpha\beta}-  \frac{\omega'-\omega}{\omega'} u_{\beta} \frac{\partial}{\partial r_{\alpha}} \nonumber\\ + \frac{1}{2}\, \frac{\omega}{\omega'} \epsilon_{\alpha\beta\gamma} (\text{rot\,} \bm u)_\gamma 
\end{align}
is the optomechanical interaction operator Eq.~\eqref{eq:L} in the coordinate representation and the polarizability operator can be readily expressed via the dielectric permittivity as  $\bm\Pi(\bm r,\omega) = \omega^2 [\bm \varepsilon(\bm r,\omega) -1]/4\pi$. When the time-inversion symmetry holds  the dielectric permittivity tensor is symmetric, so $\bm\Pi^T = \bm\Pi$. Then the distributions $\bm{\mathcal{A}}_0 (\bm r)$ and $\bm{\mathcal{A}}'^* (\bm r)$ can be calculated as the fields induced in the system by the light incident with wave vectors $\bm k$ and $-\bm k'$, respectively.

\subsection{Light scattering by a vibrating resonant layer}

We derive here the amplitudes of light scattering on a vibrating resonant layer, Eq.~\eqref{eq:huyg}.  
The layer is described by a dielectric susceptibility tensor with the components
$ \chi_{\alpha\beta}(\bm r,\omega) = \delta_{\alpha\beta} \delta(z) \chi(\omega) $  and $\chi_{zz} = \chi_{\alpha z} =\chi_{z\alpha} =0$.  Here $\alpha,\beta = x,y$ are the in-plane coordinates and $z$ is the layer normal. 
Using the Green's function
\begin{align}
D_{\alpha\beta}(\bm k,z,\omega) &= \frac{2\pi\rmi}{k_z} \left( \delta_{\alpha\beta} - \frac{k_\alpha k_\beta}{\omega^2}\right) \e^{\rmi k_z |z|} ,
\end{align}
where $\bm k = (k_x, k_y)$ is the in-plane wave vector and $k _z = \sqrt{\omega^2-\bm k^2}$,
the dressed polarization operator of the layer $\bm{\mathcal{P}} = \bm \Pi (1-\bm D \bm \Pi)^{-1}$ can be evaluated. We find
\begin{align}
\mathcal{P}_{\alpha\beta}(\bm k,z,\omega) &= \frac{\omega^2\chi(\omega)\delta(z)}{1-2\pi\rmi\omega^2\chi(\omega)/k_z} \left( \delta_{\alpha\beta} - \frac{k_\alpha k_\beta}{\bm k^2}\right)  \nonumber\\
&+ \frac{\omega^2\chi(\omega)\delta(z)}{1-2\pi\rmi k_z\chi(\omega)}  \frac{k_\alpha k_\beta}{\bm k^2} \, \,.
\end{align}
and $\mathcal{P}_{zz} =\mathcal{P}_{\alpha z} =\mathcal{P}_{z\alpha}= 0$. 
The amplitude of coherent light reflection from the layer is given by $r = \bm e'^* \cdot (2\pi \rmi \bm{\mathcal P}/k_z) \bm e_0 $. For $s$- and $p$-polarized light we obtain
\begin{align}
r_s (\bm k,\omega)&= \frac{2\pi\rmi\omega^2\chi(\omega)/k_z}{1-2\pi\rmi\omega^2\chi(\omega)/k_z} \,,\\
r_p (\bm k,\omega)&= \frac{2\pi\rmi k_z\chi(\omega)}{1-2\pi\rmi k_z\chi(\omega) } \,.
\end{align}

To calculate the amplitudes of light scattered by a layer vibration, we use the approach described in the previous section of Methods. 
First, we calculate the vector potential distribution induced by a photon at the frequency $\omega$ incident from $z\to -\infty$ with the in-plane wave vector $\bm k$ and the polarization $\bm e$,
\begin{align}
\bm{\mathcal{A}}_{\omega,\bm k} (\bm r) &= \sqrt{\frac{2\pi}{\omega}}  \e^{\rmi \bm k \cdot \bm r} \Big\{ \bm e\, \e^{\rmi k_z z} + \big[  r_s(\bm k,\omega)\,  \bm e_t  \nonumber\\
&+ r_p(\bm k,\omega)\, \bm e_l + \text{sign\,}z\, r_p(\bm k,\omega)\, \bm e_z \big]\, \e^{\rmi k_z |z|}   \Big\} \,,
\end{align}
where $\bm e = \bm e_l + \bm e_t + \bm e_z$ with $\bm e_z \parallel z$ being the out-of-plane component of the light polarization vector $\bm e$, and $\bm e_l \parallel \bm k$, $\bm e_t \perp \bm k$ being its in-plane components. 
The layer polarization induced by the photon reads 
\begin{align}\label{eq:S:PA}
\bm\Pi(\bm r, \omega) \bm{\mathcal{A}}_{\omega,\bm k} (\bm r) =  -\frac{\rmi k_z \delta(z)}{\sqrt{2\pi\omega}} \left[ r_s(\bm k,\omega) \bm e_t  + \frac{\omega^2}{k_z^2}r_p(\bm k,\omega)  \bm e_l   \right]  \e^{\rmi \bm k \cdot \bm r}.
\end{align}
Then we calculate $\bm\Lambda(\omega',\omega)\bm{\mathcal{A}}_{\omega,\bm k} (\bm r)$. Keeping in mind that according to Eq.~\eqref{eq:S:S} the result is to be multiplied by $\bm\Pi(\bm r, \omega) \bm{\mathcal{A}}_{\omega,\bm k} (\bm r)$, only needed are the in-plane components of $\bm\Lambda(\omega',\omega)\bm{\mathcal{A}}_{\omega,\bm k} (\bm r)$ at $z=0$. They read
\begin{align}\label{eq:S:LA}
\bm\Lambda(\omega',\omega)\bm{\mathcal{A}}_{\omega,\bm k}  =&\rmi \sqrt{\frac{2\pi}{\omega}} \left[ k_z  (\bm e_t + \bm e_l)  + \frac{\Omega}{\omega'} \frac{\bm k^2}{k_z} \bm e_l \right. \nonumber \\ 
&\left. - \frac{\omega}{\omega'}\frac{\bm e_l \cdot \bm k}{k_z} \bm q+ (\ldots)\, \bm e_z  \right] u_z \e^{\rmi (\bm k+\bm q) \cdot \bm r} \,,
\end{align}
where the ellipsis replaces the out-of-plane component.
Note that when evaluating the action of the optomechanical interaction operator Eq.~\eqref{eq:S:L}, in the last term describing the layer rotation we used $\text{rot\,} \bm u = 2 (\partial u_z/\partial y, -\partial u_z/\partial x,0)$. The factor 2 arises from the contribution of $\partial u_{\alpha} /\partial z$ ($\alpha=x,y$) that are non-zero even though $u_{\alpha} = 0$ at $z=0$.

Finally, we calculate the backward and forward photon scattering amplitudes
\begin{align}
R_{\rightarrow (\hookleftarrow)} = \rmi \frac{\omega'}{k_z'}u_z \int \big[ \bm\Pi \bm{\mathcal{A}}_{\omega',-\bm k'} \cdot &\bm\Lambda(\omega',\omega) \bm{\mathcal{A}}_{\omega,\bm k} \nonumber\\
\mp \bm\Lambda(-\omega,-\omega') \bm{\mathcal{A}}_{\omega',-\bm k} \cdot &\bm\Pi \bm{\mathcal{A}}_{\omega,\bm k} \big] dz\,.
\end{align}
We make use of Eqs.~\eqref{eq:S:PA} and~\eqref{eq:S:LA} and take into account that for the photon incident from $z\to +\infty$, the in-plane components of $\bm\Lambda \bm{\mathcal{A}}$ have opposite signs, while $\bm\Pi \bm{\mathcal{A}}$ is the same. Finally we obtain 
\begin{align}
&R_{\rightarrow (\hookleftarrow)} = \rmi \sqrt{\frac{\omega'}{\omega}} u_z \\
&\left\{ \left[ r_s' \bm e_t'  + \frac{\omega'^2}{k_z'^2}r_p'  \bm e_l'   \right] \cdot \left[ k_z  (\bm e_t + \bm e_l)  +  \frac{\bm e_l \cdot \bm k}{k_z} \left( \bm k - \frac{\omega}{\omega'} \bm k' \right)  \right] \right. \nonumber\\\nonumber
&\left. \pm \frac{k_z}{k_z'}  \left[ k_z'  (\bm e_t' + \bm e_l')   + \frac{\bm e_l' \cdot \bm k'}{k_z'} \left( \bm k' -\frac{\omega'}{\omega} \bm k \right)  \right] \cdot \left[ r_s \bm e_t  + \frac{\omega^2}{k_z^2}r_p  \bm e_l   \right]  \right\} ,
\end{align}
where the quantities without (with) prime refer to the initial (scattered) photon. Multiplying the result by the factor $\sqrt{\omega'/\omega}$ to relate electric fields rather than the photon amplitudes and considering $\bm e$ and $\bm e'$ that represent $s$ and $p$ polarizations, we recover Eq.~\eqref{eq:huyg} of the main text.

%%%%%%%%%%%%%%%%%%%%%%%%%%%%%%%%%%%%%%%%%%
%\nocite{apsrev41Control}
%\bibliographystyle{apsrev4}
%%\bibliographystyle{naturemag} 
%\bibliography{titleon,optokerk}

%%%%%%%%%%%%%%%%%%%%%%%%%%%%%%%%%%%%%%%%%

%merlin.mbs apsrev4-1.bst 2010-07-25 4.21a (PWD, AO, DPC) hacked
%Control: key (0)
%Control: author (8) initials jnrlst
%Control: editor formatted (1) identically to author
%Control: production of article title (0) allowed
%Control: page (1) range
%Control: year (0) verbatim
%Control: production of eprint (0) enabled
%

%%%%%%%%%%%%%%%%%%%%%%%%%%%%%%%%%%%%%%%%%

\section{Acknowledgments} 
The authors acknowledge fruitful discussions with   I.D. Avdeev, N.A. Gippius, A. Krasnok, C.R. Simovski,  A. Slobozhanyuk, and S.G. Tikhodeev. 
This work was supported by
the RFBR project 18-32-00486 and the Foundation ``Basis.'' A.N.P. and A.V.P.
also acknowledge support by the Russian President Grants No.
MD-5791.2018 and SP-2912.2016.5, respectively.

\section{Author contributions} 

A.V.P. developed the theoretical model, both A.V.P. and A.N.P. contributed to the discussion of the results and writing the manuscript. 

%%%%%%%%%%%%%%%%%%%%%%%%%%%%%%%%%%%%%%%%%
\end{document}